\documentclass[aip, preprint,  preprintnumbers]{revtex4-2}
\usepackage[a4paper,left=1.5cm,right=1.5cm,top=1.5cm,bottom=1.5cm]{geometry}
\usepackage{graphicx}
\usepackage{dcolumn}
\usepackage{bm}
\usepackage{amsmath}
\usepackage[colorlinks=true, linkcolor=blue, urlcolor=blue, citecolor=blue]{hyperref}
\usepackage[utf8]{inputenc}
\usepackage[T1]{fontenc}
\usepackage{mathptmx}
\usepackage{etoolbox}
\usepackage{float}
\usepackage{orcidlink}
\makeatletter
\def\@email#1#2{%
 \endgroup
 \patchcmd{\titleblock@produce}
  {\frontmatter@RRAPformat}
  {\frontmatter@RRAPformat{\produce@RRAP{*#1\href{mailto:#2}{#2}}}\frontmatter@RRAPformat}
  {}{}
}%
\makeatother

\begin{document}

\title{Landau-like formalism for the thermal-runaway and filamentation instabilities in switching Mott devices}

\author{Aniket Bajaj\,\orcidlink{0009-0006-4873-0274}}
\affiliation{Indian Institute of Science Education and Research Kolkata, Mohanpur, Nadia, West Bengal 741246, India}

\author{Anudeep Tullibilli\,\orcidlink{0009-0001-4174-1812}}
\affiliation{Centre for Nanoscience and Engineering, Indian Institute of Science, Bengaluru 560012, India}

\author{Arup Basak}
\affiliation{Centre for Nanoscience and Engineering, Indian Institute of Science, Bengaluru 560012, India}

\author{Pavan Nukala\,\orcidlink{0000-0002-3988-4224}}
\affiliation{Centre for Nanoscience and Engineering, Indian Institute of Science, Bengaluru 560012, India}

\author{Bhavtosh Bansal\,\orcidlink{0000-0002-7003-5693}}\email{bhavtosh@iiserkol.ac.in}
\affiliation{Indian Institute of Science Education and Research Kolkata, Mohanpur, Nadia, West Bengal 741246, India}
\date{\today}
\begin{abstract}
Correlated oxides can switch to a low-resistance state when exposed to a voltage, making these insulator-to-metal transition materials promising candidates for memristor technology. Switching in materials such as $\text{VO}_2$ occurs because of a thermal-runaway instability that leads to the abrupt formation of a hot current-carrying metallic filament within the insulator. We study, theoretically and experimentally,  this bias-induced switching as a non-linear dynamics stability problem. We construct Landau free energy-like Lyapunov functions to show that both the on-state (``{set}'') and the off-state (``{reset}'') transitions can be described as separate saddle-node bifurcations. Starting from the reaction-diffusion equation for the heat transport, the abrupt snap-back (reset) to the insulating state is systematically formulated as a mean field Landau-like dynamics with the filament width playing the role of the order parameter. The formulation predicts an abrupt first-order phase transition with a break-off of a finite width filament across the current-controlled reset transition. The formulation is quantitatively compared against our electrical and optical microscopy experiments on $\text{VO}_2$ thin film devices. Multi-cycle switching dynamics experiments further reveal the emergence of a stable stochastic regime following an initial electro-spatial forming phase. Beyond establishing a cross-disciplinary dynamical framework that may be readily compared to instabilities in various reaction-diffusion scenarios, the work also provides a predictive foundation for the rational design and optimization of Mott-transition-based technologies. 
\end{abstract}


\maketitle

The insulator–metal transition (IMT) in strongly correlated Mott materials is being actively studied for use as the switching mechanism in the next-generation nanoelectronics, neuromorphic computing, and ultrafast memory.\cite{Yang2011,Schofield2023,Lee2024,Zhang2024,Woo2025Heat} In prototypical oxides such as vanadium dioxide (VO$_2$), the phase transition can be triggered by a wide range of stimuli, including electrical or optical excitation, chemical doping, temperature, pressure or strain variation.\cite{Murakami2025,Ke2018,Kitou2026,Athermal_VO2_2025} Among these, electrical triggering is the most practical and technologically relevant approach for device implementation.\cite{Yang2013,Brockman2014,delValle2019,Pickett2012,delValle2021Science} Such electrically driven devices can exhibit ultrafast switching \cite{Brockman2014,Pickett2012,Pofelski2026} and excellent scalability.\cite{Pickett2013,SamizadehNikoo2022}

After the initial controversy regarding the nature of such bias-dependent IMT in VO$_2$, joule heating \cite{Gopalakrishnan2009,Zimmers2013,SKumar2013,Kumar2017,Polozov2020,Woo2025Heat} is now the established mechanism, rather than the nonthermal electric-field-driven breakdown.\cite{Stefanovich2000,Diener2018,Stoliar2014,Kalcheim2020} While such breakdown phenomena have been extensively addressed in the context of negative differential resistance (NDR) and nonlinear carrier dynamics in semiconductors,\cite{Ridley1963, Shaw1979,Shaw1992,Kroll1974,Kim2026} the sharp IMT intrinsic to the Mott materials leads to distinct differences and need for a new formalism. Establishing a predictive dynamical framework for the switching is vital for optimizing VO$_2$ based optoelectronic switches,\cite{Wang2025,Feng2025}, artificial neuristors \cite{Yi2018,Zhang2024,Schofield2023,Khandelwal2024} probabilistic bits,\cite{delValle2022} and stochastic computing elements.\cite{Cheng2021,Deng2023} While functional devices are widely reported, fundamental physical questions persist regarding the dynamic stability of volatile {\em set} and {\em reset} mechanism, resulting conductive filaments, absolute minimum size constraints and multi-cycle spatio-temporal evolution.

In the past few years, moving beyond empirical qualitative descriptions of the switching, there have been many attempts toward a more rigorous non-linear dynamical framework that explicitly formulates non-equilibrium potential landscapes, dynamic bifurcations, and circuit-constrained phase kinetics.\cite{Gibson2018, KumarNature2017,Brown2022,Bradicich2023,Woo2025,Kumar2018Domains,Li2017,Zhong2011,Sarles2021,Joushaghani2014,Yoon2016,Goodwill2019,Palin2025,Schneble2024,UKhandelwal2023} There have also been many previous attempts to construct Landau-like functions both for the temperature-dependent\cite{Kotliar2000,Gu2010,Shi2017,Gu2022,Zhang2025JAP} and field-induced transitions\cite{delValle2019,Shi2019,Diaz2023,Rocco2022,Shi2018PRM} to highlight the generality of the observations. But these have essentially been heuristically formulated empirical descriptions.  

Extending and generalizing the previous work, in this Letter we address, both experimentally and theoretically, the problem of switching instability in $\text{VO}_2$ channels as a catastrophic bifurcation problem.\cite{Gilmore1981} 
Starting from the well-established reaction-diffusion framework, we construct an 
an effective Landau-Ginzburg dynamics\cite{Gilmore1981, Chaikin1995} and rigorously show that the well-established condition for the joule-heating induced thermal runaway (``{\em set}'') is a saddle-node bifurcation\cite{Strogatz2024}, or equivalently a fold catastrophe, of the Lyapunov potential\cite{Wio}.  We also study the abrupt pinch-off of the metallic filament (``{\em reset}'') back to the homogeneous insulating state. The analysis of the reaction-diffusion equation and its coupling to the external circuit load lines reveals an lower bound on filament width, below which it abruptly disappears. The reset is also formulated  as phase transition problem with the filament width playing the role of an order parameter. $I(V)$ measurements synchronized with spatial optical imaging across numerous cycles quantitatively confirm these theoretical bounds.

\begin{figure}
    \centering
    \includegraphics[scale=0.15]{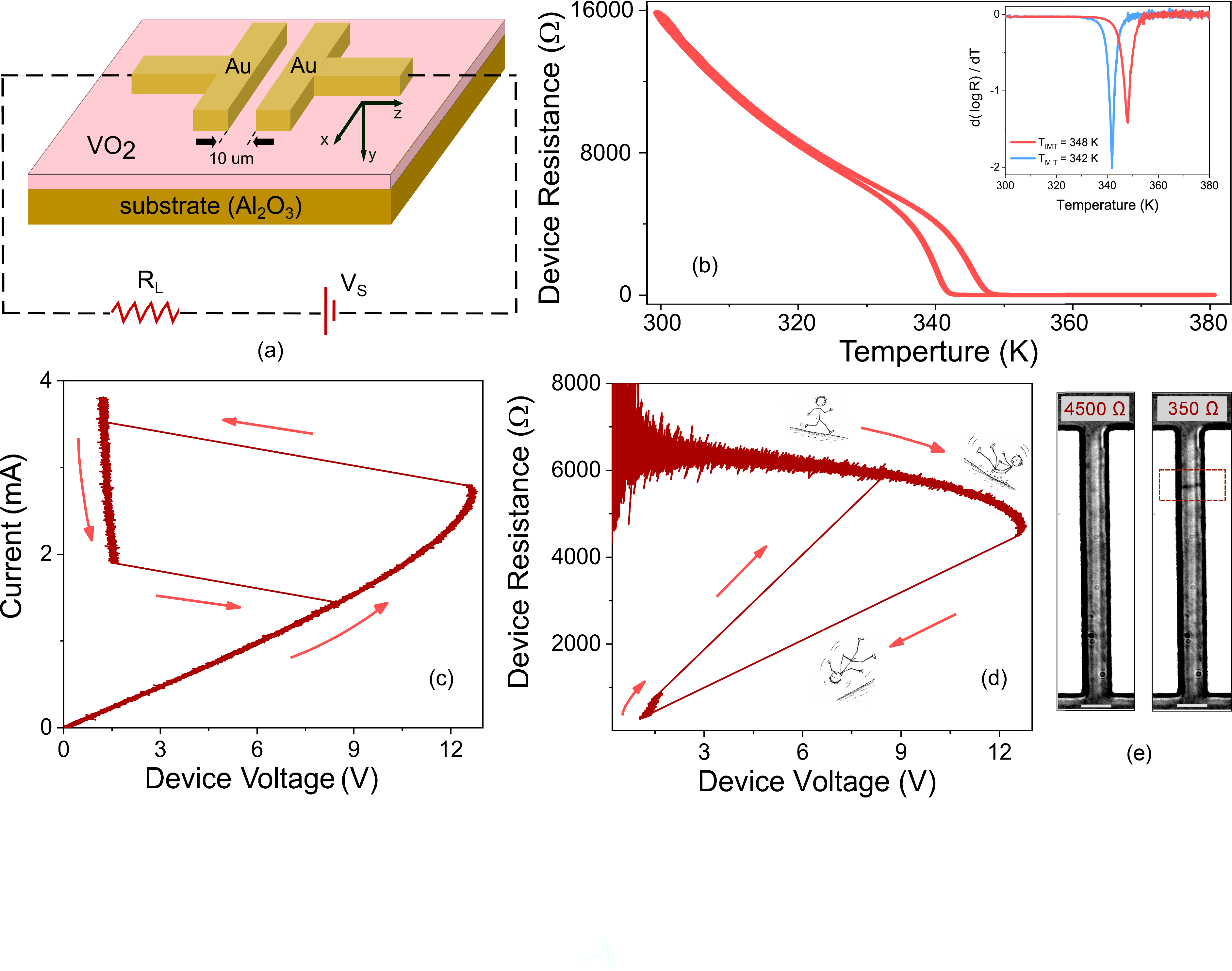}
    \caption{{\em Voltage-driven electric breakdown in VO$_2$}. (a) A two-probe VO$_2$ thin film device ($10\text{x}100$ $\mu$m) connected to a voltage source $V_s$ via a load resistor $R_L$. (b) The resistance-temperature curve of the device exhibits a negative temperature coefficient of resistance below 340 K followed by an abrupt insulator–metal transition. The transition exhibits pronounced hysteresis with $T_{IMT}\approx$348 K and $T_{MIT}\approx$342 K (inset). (c) Current–voltage $I(V)$ characteristics at 330 K under a voltage ramp show an initial linear regime followed by a pronounced nonlinear response and abrupt insulator-metal transition at $V_{IMT} =12.65$ $V$ ({\em set}) and on ramping down the voltage, the system ({\em reset}s) back to an homogeneous insulator  at $V_{MIT} =1.53$ $V$. (d) Corresponding device resistance as a function of device voltage, revealing a gradual resistance decrease, followed by a sudden resistance collapse from $4500$ $\Omega$ to $350$ $\Omega$ at the threshold voltage with appearence of metallic filament. The schematic illustrates a physical analogy of the thermal runaway process. (e) Optical images of the device channel before and after breakdown, showing the formation of a localized conductive filament bridging the electrode gap (scale bar shown is 10 $\mu$m long). }
    \label{fig:runaway}
\end{figure}

Voltage-driven breakdown was studied via transport measurements simultaneous with optical imaging, performed $\text{VO}_2$ thin film devices fabricated with gold electrodes on a sapphire substrate [Fig. \ref{fig:runaway}(a); details in Supplementary Material] using the set up described in Ref. \onlinecite{Bajaj2026}. The data persented here are on a device of $100\text{ }\mu\text{m}$ width and $10\text{ }\mu\text{m}$ gap.  Fig. \ref{fig:runaway}(b) depicts the temperature-dependent resistance $R(T)$ of the device, showing the typical negative temperature coefficient of resistance (NTCR) up to $\approx 340\text{ K}$ followed by a three-order-of-magnitude drop in $R$ at the insulator–metal transition [$T_{\text{IMT}} = 348\text{ K}$, Fig. \ref{fig:runaway}(b) inset].

Fig. \ref{fig:runaway}(c) presents current–voltage $I(V)$ characteristics measured at $330\text{ K}$ using a function generator as a voltage source ($6\text{ V/s}$ ramp) with a current-limiting series resistor ($R_L$) to prevent permanent damage [Fig. \ref{fig:runaway}(a), details of experimental setup are in Supplementary Material]. As it is well-known, enhanced joule heating at high bias combined with the NTCR response initiates self-amplifying positive thermal feedback. At a critical threshold voltage ($V_{\text{th}} = 12.65\text{ V}$), this thermal runaway instability drives an abrupt insulator-to-metal transition, causing resistance to drop from $\sim 4500\ \Omega$ to $350\ \Omega$ [Fig. \ref{fig:runaway}(d)] and the device voltage to abruptly snap from $12.65\text{ V}$ to $1.24\text{ V}$. Simultaneously performed optical imaging confirms the formation of a conductive metallic filament bridging the gap [Fig. \ref{fig:runaway}(e), dashed box].

To describe the nonlinear dynamics of the current filamentation driven by heating-induced thermal runaway instability, consider a thin-film VO$_2$ device of length $L$, width $w$ and thickness $d$ on a substrate experiencing joule heating under a local current density $\vec{j}$. The evolution of the temperature field $T(\vec{r}, t)$ within the device can be described by the reaction-diffusion equation  
\begin{equation}
c_v\frac{\partial }{\partial t}T(\vec{r},t)=\vec{\nabla}\cdot[\kappa \vec{\nabla} T]+\vec{j}\cdot\vec{E}.
\label{Eqn:Reaction-diffusion_a}
\end{equation}
Here $c_v$ is the volumetric heat capacity and $\kappa$ the in-plane thermal conductivity. Note that while the IV characteristics in [Fig. \ref{fig:runaway}(c)] is nonlinear, this is on the account of the  very strongly temperature-dependent resistivity $\rho(T)$  with $\partial \rho/\partial T<0$ [Fig. \ref{fig:runaway}(b)]. The operation is still under linear response via the Ohm's law, $\vec{E}=\rho(T)\vec{j}$. Integrating over film thickness $d$ along the $y$ direction converts interface heat loss into an effective volumetric sink,  reducing one dimension in the reaction-diffusion equation above but at the cost of introducing the Newton cooling term $\frac{h}{d}(T-T_0)$ for the heat conduction to the substrate at the temperature $T_0$. $h$ is the surface heat transfer coefficient and $d$ is film thickness. Eq. \ref{Eqn:Reaction-diffusion_a} thus reduces to
\begin{equation}
c_v\frac{\partial }{\partial t}T(\vec{x},t)=\kappa \nabla_\parallel^2 T(\vec{x},t)+\frac{E^2}{\rho(T)}-\frac{h}{d}(T-T_0).
\label{Eqn:Reaction-diffusion}
\end{equation}
Here $\vec{x}$ is a two-dimensional vector in the xz-plane with $z \parallel$ current flow and $\nabla_\parallel^2 = \frac{\partial^2}{\partial x^2} + \frac{\partial^2}{\partial z^2}$. 
\begin{figure}[t!]
    \centering
    \includegraphics[scale=0.19]{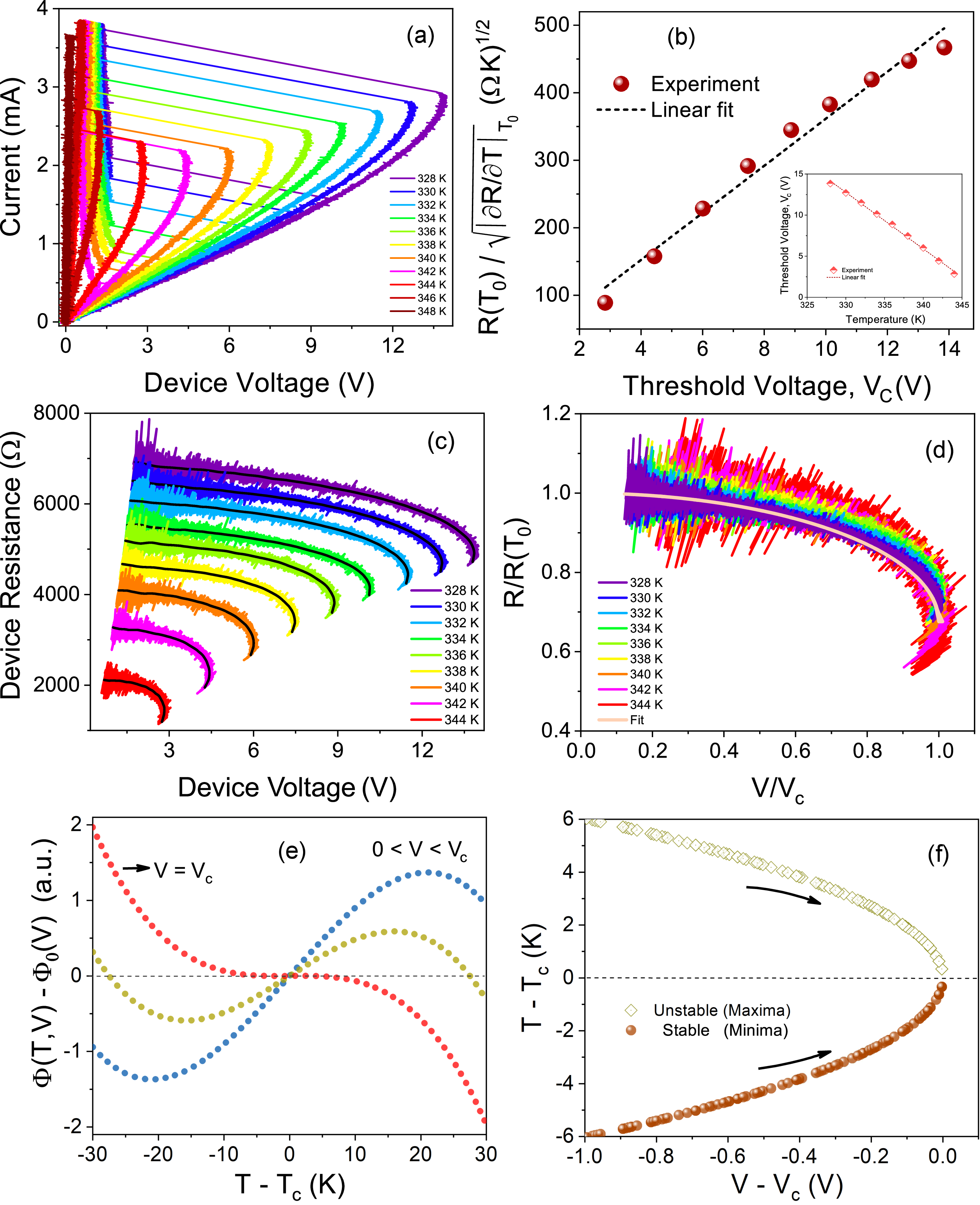}
    \caption{{\em Experimental evidence of thermal runaway instability.} (a) Temperature variation of I(V) characteristics. Hysteretic behaviour transforms to linear behaviour as T approaches 348 K ($T_{IMT}$)  (b) Threshold voltage ($V_c$) decreases linearly with increasing temperature, highlighting the thermal origin of the breakdown. (c) Threshold voltage ($V_c$) plotted against $R(T_0)/\sqrt{|\partial R/\partial T|_{T_0}}$ at different temperatures, showing a linear relationship consistent with the predicted thermal runaway condition. (c) Device resistance curves plotted as a function of device voltage at different temperatures (Smoothed data is represented by the overlaid black curves). (d) shows the collapse of resistance curves at different temperatures by rescaling the device resistance  with equilibrium resistance; [$R/R(T_0)$] and device voltage with threshold voltage; $(V/V_c)$. (e) Lyapunov potential $\Phi(T,V)$ illustrating the evolution of the saddle-node bifurcation\cite{Strogatz2024} as applied voltage approaches threshold voltage ($V\rightarrow V_c$). (f) Bifurcation diagram of the steady-state temperature solutions, showing the merging of stable and unstable branches at the threshold voltage. }
    \label{fig:temp_IVrunaway}
\end{figure}

Let us now write a zero-dimensional ``mean-field'' lumped-circuit version of Eq. \ref{Eqn:Reaction-diffusion} by integrating it over the thin film volume and assuming a uniform temperature over the film 
\begin{equation}
\frac{dT}{dt}
= \frac{1}{C_{th}}\left[\frac{V^2}{R(T)} - K_{th} (T - T_0)\right]=f(T,V). 
\label{lumpedckteqn}
\end{equation}
The uniform temperature assumption is a good approximation before the critical field, beyond which it completely breaks down as filamentation sets in. Here we have used $\rho=\frac{A}{L}R$ and $K_{th}=\frac{hAL}{d}$  as total thermal conductance with $A$ is the area of cross section of device channel. $V=E/L$, where $L$ is the distance between the contacts. 

 Writing $R(T)\simeq R(T_0)-|\frac{\partial R}{\partial T}|_{T_0}[T-T_0]$, the steady state solution of Eq. \ref{lumpedckteqn} ($\partial T/\partial t=0$) yields the following expression for the temperature rise $\Delta T$ during steady-state device operation:
 \begin{equation}
\Delta T={T-T_0\simeq \frac{R(T_0) + \sqrt{R^2(T_0) - \frac{1}{K_{th}}|\frac{\partial R}{\partial T}|_{T_0} V^2}}
{2 |\frac{\partial R}{\partial T}|_{T_0}}}.
\label{Temp_roots}
\end{equation}
The solution is meaningful as long as $R^2(T_0)>\frac{4 }{K_{th}}|\frac{\partial R}{\partial T}|_{T_0} V^2$. Violation of this inequality yields a thermal runaway instability, with the critical voltage $V_c$ given by
\begin{equation} {V^2_c=\frac{K_{th}}{4} \frac{R^2(T_0)}{|{\partial R}/{\partial T}|_{_{T_0}}}}.
\label{Vc_critical_voltage}
\end{equation}
The discussion so far is builds up to what is already well-known and standard\cite{Zimmers2013,SKumar2013,Bradicich2023,Polozov2020,Kumar2018Domains,Gibson2018} but it is interesting to attempt an experimental validation of Eq. \ref{Vc_critical_voltage}. This is done in Fig. \ref{fig:temp_IVrunaway}(a, b). As the ambient temperature increases from $328\text{ K}$ to $ 348\text{ K}\ (T_{\text{IMT}})$, the $I(V)$ hysteresis gradually diminishes, following to linear Ohmic behavior at $T_{\text{IMT}}$ accompanied by a progressive drop\cite{AdvElectronMater2016} in threshold voltage $V_c$ [Fig. \ref{fig:temp_IVrunaway}(b) inset]. Extracting $R(T_0)$ from the low-bias regime reveals that $V_c$ indeed seems to scale linearly with $R(T_0)/\sqrt{\vert{}\partial R/\partial T\vert{}_{T_0}}$ across temperatures [Fig. \ref{fig:temp_IVrunaway}(b)].

Fig. \ref{fig:temp_IVrunaway}(c) presents the corresponding bias-dependent resistance from the data in Fig. \ref{fig:temp_IVrunaway}(a) up to the switching threshold. All these curves nearly collapse onto each other, when the x and y axes are scaled to $V/V_c$ and $R/R(T_0)$ respectively Fig. \ref{fig:temp_IVrunaway}(d) and seem to fit the semi-empirical [See Supplement Material] relationship 
\begin{equation}
\frac{R(T)}{R(T_0)} = \frac{R(T_c)}{R(T_0)} \pm \left[ 1-\frac{R(T_c)}{R(T_0)} \right ] \sqrt{1 - \left(\frac{V}{V_c}\right)^2}.
\label{eq:master_scaling}
\end{equation}
Note that the scaling behavior would be different for field-driven electronic mechanisms (e.g., Poole–Frenkel emission or impact ionization).\cite{Diener2018}

While the discussion so far is standard, it is perhaps useful to rigorously contextualize the runaway instability within the framework non-equilibrium phase transitions and catastrophe theory. 

\noindent
{\it Lyapunov stability}---Eq. \ref{lumpedckteqn} is a one-dimensional equation and we can always write $f(T,V)=-\frac{d}{dT}\Phi(T,V)$ 
with 
\begin{equation}
\Phi(T,V)=\frac{1}{C_{th}}\left[\frac{K_{th}}{2}(T-T_0)^2-V^2\int_{T_0}^T\frac{1}{R(T')}dT'\right].
\label{Eqn:Lyapunov functional}
\end{equation}
Eq. \ref{lumpedckteqn} thus describes a fully dissipative gradient flow   (Model A)\cite{Chaikin1995} dynamics
\begin{equation}
\frac{\partial T}{\partial t}=-\frac{d}{dT}\Phi(T,V),
\label{eqn:RelaxDynamics}
\end{equation}
and $\Phi(T,V)$ is a Lyapunov function\cite{Wio} since 
\begin{equation}
\frac{d}{dt}\Phi(T,V)=\frac{d}{dT}\Phi(T,V)\frac{dT}{dt}=-\left[\frac{d\Phi}{dT}\right]^2\leq 0. 
\label{eqn:LyapunovStability1}
\end{equation}
The system should thus evolve to the minimum of 
$\Phi(T,V)$ in steady state, making $\Phi(T,V)$  analogous to a mean-field Landau free energy\cite{Chaikin1995} (even though it has units of Watt-Kelvin) with the device temperature $T$ being the order parameter. The two terms in Eq. \ref{Eqn:Lyapunov functional} correspond to the cost of the temperature excursion against the heat sink and the electrical work accumulated over this excursion that is made available by the field.

\begin{figure}[t!]
    \centering
    \includegraphics[scale=0.19]{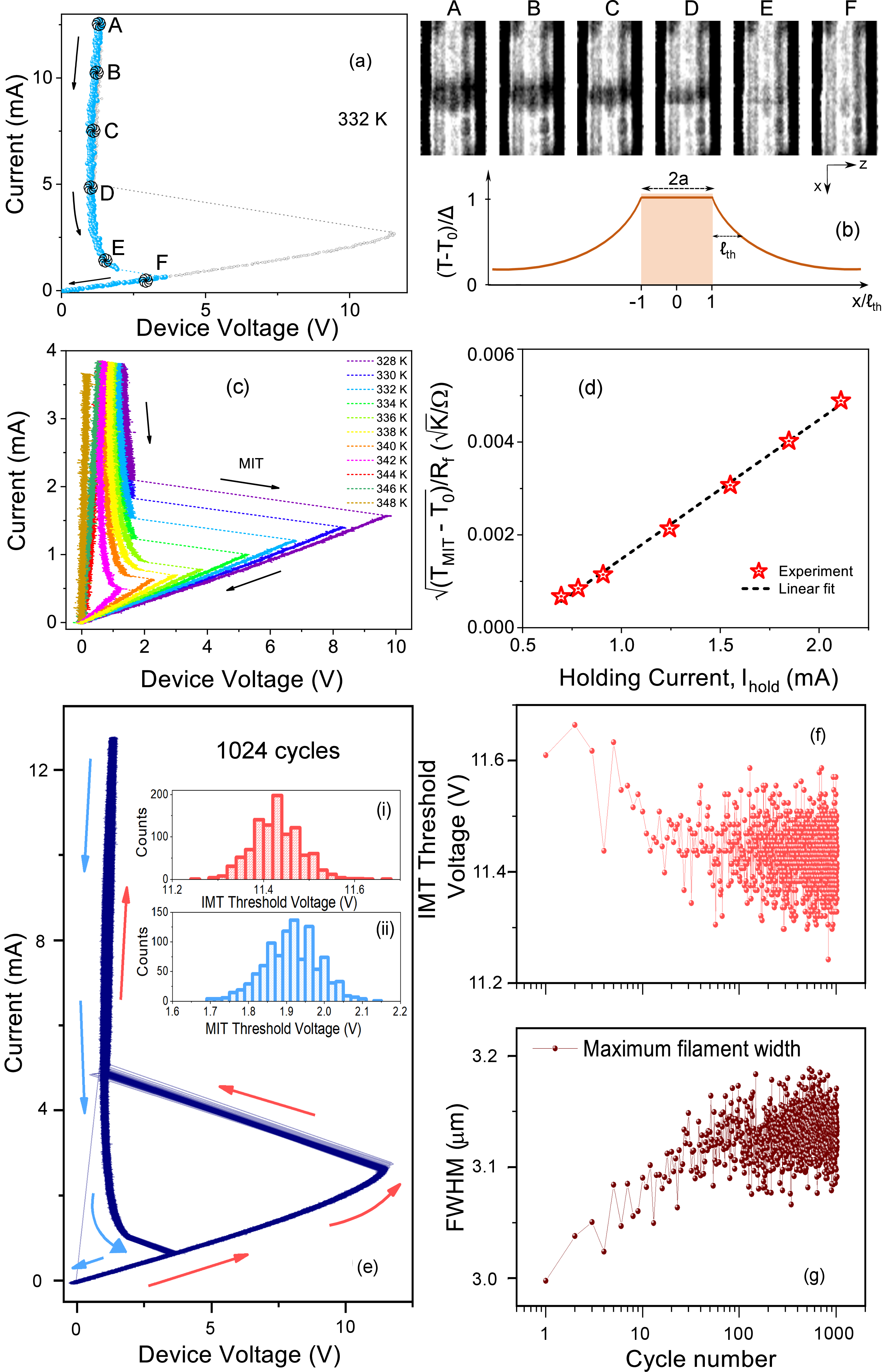}
    \caption{{\em Shrinking of the metallic filament up to the {\em reset} instability.} (a, b) $I(V)$ characteristics at 332 K alongside corresponding close-up optical images of a $10 \times 100\,\mu\text{m}^2$ device. (b) Filament's temperature profile
    assumed to be central metallic core (shaded region) ($|x| \le a$) held at $T_c$ and flanked by exponential tails in the insulating region ($|x| > a$) decaying to $T_0$. (c) Temperature-dependent $I(V)$ sweeps highlighting the {\em Reset} branch [extracted from Fig. \ref{fig:temp_IVrunaway}(a)]. (d) Linear scaling of $I_{\text{hold}}$ versus $\sqrt{T_{\text{MIT}} - T_0}/R_f$ (dashed fit), confirming the thermal stability condition. (e) Reversible $I(V)$ switching across 1024 cycles at 332 K, with insets showing statistical distributions for (i) the IMT threshold voltage and (ii) the MIT threshold voltage. (f) Evolution of the IMT threshold voltage with cycle number (log scale), revealing an initial forming phase prior to stabilization. (g) Corresponding maximum filament width (FWHM) per cycle, demonstrating an inverse correlation with the electrical trend in (f) prior to achieving long-term spatial stability.}
    \label{fig:reset_process}
\end{figure}

The thermal runaway instability at $(T_c, V_c)$ is a critical point of $\Phi(T,V)$. This is easily verified by noting that $\Phi^c_T\equiv \frac{\partial}{\partial T} \Phi(T,V)|_{T_c,V_c}=0$ and $\Phi^c_{TT}\equiv \frac{\partial^2}{\partial T^2} \Phi(T,V)|_{T_c,V_c}=0$.
Here $V_c$ given by Eq. \ref{Vc_critical_voltage} and $T_c =  T_0 +\frac{R(T_0)}{{2|{\partial R}/{\partial T}|_{_{T_0}}}}$. Taylor-expanding $\Phi(T,V)$ up to third order in $(T - T_c)$ and first order in $(V - V_c)$ around the critical point 
yields the normal form of the saddle-node bifurcation\cite{Strogatz2024} (or equivalently, a fold catastrophe\cite{Gilmore1981})
\begin{eqnarray}
\Phi(T, V) \approx \Phi(T_c,V_c)  
+ \Phi^c_{TV}(V - V_c)(T - T_c) \nonumber 
+ \frac{1}{6}\Phi^c_{TTT}(T - T_c)^3. 
\label{Eqn: Lypunov-Normal}
\end{eqnarray}
Substituting in Eq. \ref{eqn:RelaxDynamics}, we get 
\begin{equation}
\frac{dT}{dt} = \frac{4 V_c}{R(T_0)} (V - V_c) +  \frac{2 K_{th}}{R(T_0)}{\left|\frac{\partial R}{\partial T}\right|_{_{T_0}} (T - T_c)^2}.
\label{T_dot}
\end{equation}
Here the coefficients of $(V-V_c)(T-T_c)$ and $(T-T_c)^3$ correspond, respectively, to the $\Phi^c_{TV}$  and $\Phi^c_{TTT}$ in Eq. \ref{Eqn: Lypunov-Normal} with taking $C_{th}$ = 1. The two steady-state solutions ($\partial T/\partial t = 0$) of Eq. \ref{T_dot},  
$T^* = T_c \pm  \left[{R(T_0)}/\left| \frac{\partial R}{\partial T} \right|_{T_0}\right]\sqrt{\frac{1}{2} \left(1-\frac{ V}{V_c} \right)}$, 
are shown in Fig. \ref{fig:temp_IVrunaway}(f) and correspond to the minimum (stable) and the maximum (unstable) of the Lyapunov function [Fig. \ref{fig:temp_IVrunaway}(e)], respectively. As $V \to V_c$, these fixed points coalesce at $T_c$ and annihilate via a saddle-node bifurcation [Fig. \ref{fig:temp_IVrunaway}(e)]. For $V > V_c$, real equilibrium solutions cease to exist leading to a thermal runaway. A hot metallic filament of finite half-width $a$ abruptly nucleates within the insulating matrix resulting in a dramatic decrease in the device resistance. This results in a transition from voltage-controlled to current-controlled device operation, with the current $I\simeq V_s/R_L$  now limited by the $R_L$ the load resistance in the circuit [Fig. \ref{fig:runaway}(a)]. $V_s$ is the source voltage. The metallic filament grows with increase in the  current. With the decrease in the current the filament width shrinks to a minimum half-width $a_\textrm{min}$, when it abruptly disappears causing a {\it reset} back to the homogeneous insulating state [Fig. \ref{fig:reset_process}(a, b)].

Let us now formulate a spatiotemporal description of this {\em reset} as a non-equilibrium phase transition, where the metallic filament's half-width $a(t)$ plays the role of an order parameter that abruptly goes to zero at the reset instability. 

We again start with Eq. \ref{Eqn:Reaction-diffusion} but which now should also include the latent heat of transformation since filamentation involves  nucleation of the metallic phase. This is accomplished by replacing $c_v(\partial T/\partial t)\rightarrow\partial H/\partial t$ where the volumetric enthalpy density\cite{Kumar2018Domains}  $H(T) = c_v (T - T_0) + \varrho L_h \Theta(T - T_c)$, $c_v$ is the volumetric heat capacity, and $\Theta(T - T_c)$ is the Heaviside step function enforcing latent heat $\varrho L_h$ absorption at $T_c$. Eq. \ref{Eqn:Reaction-diffusion} may thus be written as 
\begin{equation}
d\frac{\partial H(T)}{\partial t} = d\kappa \nabla_\parallel^2 T(\vec{x},t) + d\frac{E^2}{\rho(T)} - h(T - T_0),
\label{Eqn:1D_modified_RD}
\end{equation}
In what follows, we use a series of approximations to convert Eq. \ref{Eqn:1D_modified_RD} into an ordinary nonlinear differential equation amenable to catastrophe theory\cite{Gilmore1981, Strogatz2024} analysis. This is accomplished by first eliminating the z-dependence by assuming thermal uniformity ($\partial^2 T/\partial z^2 = 0$) along the current flow. We shall now employ a simplified version of the Gelarkin projection method\cite{Fletcher-Galerkin} by next making the following ansatz for the lateral temperature profile  [Fig. \ref{fig:reset_process}(b)] 
\begin{equation}
\hat{T}(x; a) - T_0 = \begin{cases} 
\Delta, & |x| \le a \\ 
\Delta \exp\left(-\frac{|x| - a}{\ell_{\text{th}}}\right), & |x| > a. 
\end{cases}
\label{Eqn:ProfileAnsatz}
\end{equation}
Here $\Delta=T_c-T_0$ and $\ell_{\text{th}} = \sqrt{\kappa d / h}$ is the thermal healing length. For $|x|-a\gg \ell_{\text{th}}$ the channel is at the substrate temperature $T_0$; it is assumed that the current flows only through the metallic filament. We now integrate over the x coordinate to be left with only a 1D moving-boundary ordinary differential equation in the variable $a(t)$. 

The enthalpy density $H(\hat{T}(x))$ is integrated across the transverse profile $\hat{T}(x; a(t))$ to yield enthalpy per unit channel length $\mathcal{U}(a)$ (see Supplementary Material)
\begin{equation}
\frac{\partial}{\partial t}\mathcal{U}(a) \equiv d\frac{\partial}{\partial t}\int_{-\infty}^{\infty} H(\hat{T}(x)) \, \mathrm{d}x.
\label{Eqn:TotalEnthalpyDef}
\end{equation}
Since the spatiotemporal profile depends on time exclusively through the scalar phase-boundary order parameter $a(t)$, the spatial enthalpy accumulation rate into  interfacial velocity $\dot{a}(t)$ via the product rule
\begin{equation}
\frac{\mathrm{d}\mathcal{U}}{\mathrm{d}t} = \left( \frac{\mathrm{d}\mathcal{U}}{\mathrm{d}a} \right) \frac{\mathrm{d}a}{\mathrm{d}t} = \mathcal{E}_S \frac{\mathrm{d}a}{\mathrm{d}t}.
\label{Eqn:ChainRuleEnthalpy}
\end{equation}
Here $\mathcal{E}_S \equiv \frac{\mathrm{d}\mathcal{U}}{\mathrm{d}a} = 2d(c_v \Delta + \varrho L_h)$ represents the effective volumetric enthalpy cost per unit lateral area. 

Similarly integration of the right side of Eq. \ref{Eqn:1D_modified_RD}
(after substituting Eq. \ref{Eqn:ProfileAnsatz}) yields the excess power $\mathcal{P}(a, I)$ across electrode gap $L$ and Eq. \ref{Eqn:1D_modified_RD} (see Supplementary Material) can be transformed to 
\begin{equation}
\mathcal{E}_S L \frac{\mathrm{d}a}{\mathrm{d}t} = \mathcal{P}(a, I)= I^2 \left(\frac{\rho_M L}{2 a d}\right) - 2 h \Delta (a + \ell_{\text{th}})L.
\label{Eq.dynamics_reset}
\end{equation}
Since the post-filamentation current-control is achieved via the load resistance $R_L$ [Fig. \ref{fig:runaway}(a)], it is natural to write $\mathcal{P}(a, V_{\text{s}}, R_L)$ instead of $\mathcal{P}(a, I)$. With $ I= V_{\text{s}}/(R_L + \frac{\rho_M L}{2ad})$
\begin{equation}
\mathcal{P}(a, V_{\text{s}}, R_L) = \frac{V_{\text{s}}^2 \left(\frac{\rho_M L}{2ad}\right) }{\left(R_L  + \frac{\rho_M L}{2ad}\right)^2} - 2h\Delta(a + \ell_{\text{th}})L.    
\label{NetPower_Vs}
\end{equation}
The structure of Eq. \ref{Eq.dynamics_reset} implies that by defining
\begin{equation}
  \Psi(a, V_{\text{s}}, R_L) = -\frac{1}{\mathcal{E}_SL} \int_{0}^{a} \mathcal{P}(a', V_{\text{s}}, R_L) \, \mathrm{d}a',
  \label{Eq.lyuponov_reset}
\end{equation}
it can be recast as a gradient relaxation dynamics of $a(t)$
\begin{equation}
  \frac{\mathrm{d}a}{\mathrm{d}t} = -\frac{\partial }{\partial a}\Psi(a, V_{\text{s}} R_L),
\label{RelaxKinetics}
\end{equation}
and $\partial \Psi/\partial t<0$ (see Eq. \ref{eqn:RelaxDynamics} and \ref{eqn:LyapunovStability1}). Thus  $\Psi(a,V_s,R_L)$ is also a Lyapunov function. 
The equilibrium states, defined by roots $\mathcal{P}(a, V_{\text{s}}) = 0$, also correspond to the minimum $\partial \Psi / \partial a = 0$ of the ``potential'' $\Psi$. We have thus recast the kinetic balance into an order-parameter relaxation framework within an effective non-equilibrium Landau-like potential $\Psi(a, V_{\text{s}}, R_L)$ with the filament half-width $a(t) \ge 0$ acting as the  order parameter. 

The critical point $(a_{\min}, V_{s,\min})$ is determined for the minimum value of $V_s$ below which there are no real solutions for $a$ and its nature is determined 
by constructing the normal form of the bifurcation. We can do this by Taylor-expanding $\Psi(a, V_s)$ to third order in $(a - a_{\min}) \equiv \delta a$ and first order in $(V_s - V_{s,\min}) \equiv \delta V_s$ 
\begin{equation}
\Psi(a, V_s) \approx \Psi^0 + \Psi^0_{a V_s} \delta V_s \delta a + \frac{1}{6}\Psi^0_{aaa}(\delta a)^3, 
\label{Eqn:Lyapunov-Normal2}
\end{equation}
since $\Psi^0_a$=$\Psi^0_{aa}$=$0$. Here $\Psi^0_{a_1,a_2,..a_n}$ denotes $\partial^n \Psi/\partial a_1\partial a_2...\partial a_n$ evaluated at the critical point $(a_{min}, V_{s,min})$.
Substituting in Eq. \ref{RelaxKinetics} and noting that $da/dt=d\delta a/dt$, the phase-boundary motion near threshold has the normal form of the saddle-node bifurcation\cite{Strogatz2024}
\begin{equation}
\frac{\mathrm{d}\delta a}{\mathrm{d}t} =- \Psi^0_{a V_s} \delta V_s - \frac{1}{2}\Psi^0_{aaa}(\delta a)^2.
\label{Eqn:SaddleNodeNormalForm}
\end{equation}
The expansion coefficients are explicitly given by: $\Psi^0_{a V_s} =  -\frac{4h\Delta(a_{\min} + \ell_{\text{th}})}{\mathcal{E}_S V_{s,\min}} < 0$. $\Psi^0_{aaa}> 0$ requires \\ $\frac{1}{\mathcal{E}_S L} \frac{2 R_L R_{f,\min} V_{s,\min}^2 (2 R_{f,\min}-R_L)}{a_{\min}^2 (R_L + R_{f,\min})^4} > 0$ or $R_{f,min}>R_L/2$ near critical point. This condition is hard to verify experimentally because the resistance measured in Fig. \ref{fig:temp_IVrunaway}, for example, has a non-negligible contribution of the parallel resistance of the insulating region that the model ignores.  Above the threshold ($ V_s > V_{s,\min}$), the system possesses two real fixed points, a stable node $a_{min,\text{stable}} \propto -\sqrt{ V_s- V_{s,min}}$ corresponding to the conductive metallic filament branch, and an unstable threshold barrier $a_{min,\text{unstable}} \propto \sqrt{ V_s- V_{s,min}}$. At the threshold ($ V_s = V_{s,\min}$), the fixed points annihilate via a saddle-node bifurcation, leaving no local real roots for $\mathrm{d}a/\mathrm{d}t = 0$. The filament experiences deterministic thermal runaway toward contraction ($\mathrm{d}a/\mathrm{d}t < 0$), triggering volatile channel pinch-off and resetting the device to its high-resistance insulating state.

The minimum possible half-width $a_{\min}$ for the filament can be derived from  Eq. \ref{NetPower_Vs}  by dividing the stationary condition $\mathcal{P} = 0$ by the zero-curvature condition $\partial\mathcal{P}/\partial a = 0$ to yield 
\begin{equation}
\frac{a_{\min}}{\ell_{\text{th}}} = -\frac{1}{4}\pm\frac{1}{4}\left(\sqrt{1 + 8\frac{a_*}{\ell_{\text{th}}}}\right). 
\label{Eq.min_width_fila}
\end{equation}
 Here $a_* \equiv \rho_M L / (2d R_L)$. Eq. \ref{Eq.min_width_fila} also follows directly from the circuit load line, suggesting an overall consistency of the formulation. Combining the steady-state condition $\mathcal{P}(a, I) = 0$  with $ I= V_{\text{s}}/(R_L + \frac{\rho_M L}{2ad})$ we get the relationship between the circuit parameters and the filament width,  
\begin{equation}
V_{\text{s}} = 2 d R_L \sqrt{\frac{\kappa \Delta}{\rho_M} \left[\frac{a}{\ell_{\text{th}}} \left(\frac{a}{\ell_{\text{th}}}+1\right)\right]} + L \sqrt{\frac{h \Delta \rho_M}{d}\left(1+\frac{1}{\frac{a}{\ell_{\text{th}}}}\right)}.
\label{eq:reducedloadline}
\end{equation}
Absence of the load resistance in the circuit would cause an unconstrained thermal runaway and the transformation of the complete device to the metallic state.\cite{fn1}
For wide filaments ($a \gg \ell_{\text{th}}$), the first term dominates; channel widening lowers device resistance and increases current, causing $R_L$ to absorb most of $V_{\text{s}}$ and self-limit further expansion. 
Conversely, for narrow filaments ($a \ll \ell_{\text{th}}$), the second term dominates. As $a \to 0$, persistent lateral edge dissipation ($2 h \Delta \ell_{\text{th}}$) demands a divergent internal bias  ($V_{\text{dev}} \to \infty$) to maintain thermal equilibrium. Consequently, real solutions of Eq. \ref{eq:reducedloadline} exist only beyond a minimum $V_{\text{s}}$. Zero-width cores are physically unsustainable, ensuring the filament stabilizes at a finite minimum width $a_{\min}$ determined from the extremum of Eq. \ref{eq:reducedloadline}. Setting $\mathrm{d}V_{\text{s}}/\mathrm{d}(a/\ell_{\text{th}}) = 0$ yields Eq. \ref{Eq.min_width_fila}. Parenthetically note that the formalism ignores the parallel resistance of the insulating region which will become important as the $a\rightarrow 0$, but this will alter the overall qualitative picture. 

We thus establish that a metallic filament cannot exist or nucleate continuously from zero size and must have a  minimum width ($2a_{\min}$) that is reached at $V_{\text{s},\min}$ just prior to the {\em reset} switching. Since during {\em set} switching, the filament nucleates with $V_{set} > V_{\text{s},\min}$, the initial width at {\em set} exceeds $2a_{\min}$. The switching is hysteretic because the material can exist with two very different values of resistivity corresponding to the metallic and the insulating states. The small resistance of an existing metallic filament ensures that there is already enough heat being generated for the same bias (larger $I$ in the metallic state) even at voltages smaller than $V_{{set}}$.

These critical behaviors are directly validated by experimental observations (Fig. \ref{fig:reset_process}). Optical imaging [Fig. \ref{fig:reset_process} (a, b)] confirms physical filament contraction during current ramp-down ($12.7\text{ mA} \to 1.4\text{ mA}$). Below a critical holding current $I_{\text{hold}} \approx 0.6\text{ mA}$, the metallic channel abruptly pinches off, triggering a volatile {\em reset} [Note: Optical imaging was conducted independently for the $I(V)$ sweeps in Fig. \ref{fig:reset_process} (a) and (e) using a smaller series resistance to achieve higher currents].

 Elevating the ambient temperature $T_0$ toward $T_{\text{MIT}}$ systematically lowers the {\em reset} threshold [Fig. \ref{fig:reset_process}(c), these I(V) curves correspond to {\em reset} branch of curves in Fig. \ref{fig:temp_IVrunaway}(a)], driven by the intrinsic thermal factor $\sqrt{\Delta} = \sqrt{T_{\text{MIT}} - T_0}$ explicit in both terms of Eq. \ref{eq:reducedloadline}. This holding-state scaling is quantitatively confirmed in Fig. \ref{fig:reset_process}(d).  In steady state, because metallic channel resistance scales inversely with core width \\ ($R_f = \rho_M L/2ad \propto 1/a$), and for narrow filaments near the threshold ($a \sim \ell_{\text{th}}$), the holding current  scales with  $I_{\text{hold}} \propto d\sqrt{\frac{\kappa \Delta}{\rho_M}}\left(\frac{a}{\ell_{\text{th}}}\right)$ (using Eq. \ref{Eq.dynamics_reset}). This yields the observed condition $I_{\text{hold}} \propto \frac{\sqrt{T_{\text{MIT}} - T_0}}{R_f}$ shown in Fig. \ref{fig:reset_process}(d). This linear relationship confirms that the load-line dictates the minimum holding condition across temperatures. 
 
Reproducibility and stochasticity in the resistive switching behavior of VO$_2$ has also been extensively studied\cite{Roldan2023,Cheng2021,Deng2023,Bidoul2024} for its technological relevance. Fig. \ref{fig:reset_process}(e-g) extend the previous work by also simultaneously imaging the bias-dependent filament over 1024 cycles. We find an initial ``forming phase'' in Fig. \ref{fig:reset_process}(f,g) during the initial cycles which slowly evolves to a stable stochastic switching regime and a correlation between the electronic and spatial stability of the metallic filament.

During the initial 50-cycle forming window [Fig. \ref{fig:reset_process}(f)], the threshold voltage undergoes a progressive drop, while synchronized spatial imaging reveals that the metallic filament's full width half maximum (FWHM) at peak current [image A in Fig. \ref{fig:reset_process}(b)] expands across this window [Fig. \ref{fig:reset_process}(g)], inversely correlating the drop in threshold voltage. This initial structural stabilization—driven by strain accommodation and defect redistribution—widens the metallic channel to lower the critical insulator-to-metal transition threshold.\cite{Cheng2021,Roldan2023,Conti2026}

Beyond this forming regime, the threshold voltage and filament width saturate with minimal spatial fluctuations around a steady mean, establishing a long-term stochastic switching regime optimal for neuromorphic architectures.\cite{Roldan2023,Yi2018,Cheng2021,delValle2022,Deng2023,Bidoul2024} A video capturing all 1024 cycles is in the Supplementary Material.

In summary, we have studied the thermal runaway ({\em set}) and the pinch-off ({\em reset}) of the metallic filament in VO$_2$ Mott device as the loss of Lyapunov stability via saddle-node bifurcations (fold catastrophes). Starting from the reaction-diffusion equation for the heat balance, effective Landau-like mean field descriptions were formulated for both these non-equilibrium transitions. By quantitatively bridging threshold instabilities ({\em set}/{\em reset}) with spatiotemporal filament dynamics, our work yields a broad perspective on this very well-studied problem and yields  insights for rational optimization of Mott-transition technologies.


\section*{Supplementary Material}
See the Supplementary Material for details of sample and device fabrication, experimental setup descriptions, and complete derivations of the pre-threshold resistance scaling relation and moving-boundary filament dynamics. A supplementary video capturing 1024 cycles of real-time optical imaging during resistive switching is also provided.

\section*{Acknowledgments}
A.B. acknowledges the Indian Institute of Science
Education and Research Kolkata for fellowship. All the authors acknowledge Harish Kumar Maurya for his assistance with device optimization. B.B. thanks the Science and Engineering Research Board (SERB), Department of Science and Technology, Government of India, for the Core Research Grant (No. CRG/2022/008662). Large Language Models [Claude (Opus 5 and Fable 5.1), Google Gemini (3.6 Thinking)] 
were used while formulating and refining the theoretical discussion in the manuscript.

\section*{AUTHOR DECLARATIONS}
\section*{Conflict of Interest}
The authors have no conflicts to disclose.
\section*{Author Contributions:}
\textbf{Aniket Bajaj:} Conceptualization (equal); Data curation (lead); Formal analysis (lead); Investigation (equal); Methodology (equal); Resources (equal); Software (lead); Validation (equal); Visualization (equal); Writing – original draft (lead); Writing – review and editing (lead). \textbf{Anudeep Tullibilli:} Investigation (supporting); Methodology (supporting); Resources (equal). \textbf{Arup Basak:} Investigation (supporting); Methodology (supporting); Resources (equal). \textbf{Pavan Nukala:} Funding acquisition (equal); Investigation (supporting); Project administration (supporting); Resources (equal); Validation (supporting). \textbf{Bhavtosh Bansal:} Conceptualization (lead); Formal analysis (lead); Funding acquisition (lead); Project administration (lead); Resources (lead); Supervision (lead); Validation (lead); Writing – original draft (equal); Writing –review and editing (lead).

\clearpage
\setcounter{figure}{0}
\renewcommand{\thefigure}{S\arabic{figure}}

\begin{center}
    \textbf{\Large Supplementary Material}
\end{center}

This supplementary material provides information on sample preparation and device fabrication. Details of the experimental setup used for simultaneous electrical transport measurements and optical imaging are provided. Complete derivation of prethreshold resistance scaling relation and moving-boundary dynamics of filamentation are also provided.   


\maketitle

\section{Details of sample preparation and device fabrication}

Epitaxial VO$_2$ thin films were deposited on single-crystal Al$_2$O$_3$ (0001) sapphire substrates (10 × 10 × 0.5 mm$^3$) using pulsed laser deposition (PLD) with a 248 nm KrF excimer laser. Prior to deposition, the substrates were cleaned with acetone and isopropanol (IPA). A V$_2$O$_5$ target was pre-sputtered using 1500 laser pulses immediately prior to film growth to remove surface contaminants and ensure stable ablation conditions. The VO$_2$ films were deposited at a substrate temperature of 600 $^\circ$C under an oxygen partial pressure of 10$^{-2}$ mbar. The laser fluence was maintained at 1.2 Jcm$^{-2}$, with a repetition rate of 3 Hz and a pulse duration of 20 ns. The target-to-substrate distance was fixed at approximately 50 mm. Following deposition, the samples were cooled to room temperature at a rate of 30 $^\circ$C min$^{-1}$ while maintaining the same oxygen partial pressure used during growth. The resulting VO$_2$ films had a thickness of approximately 65 nm.

\begin{figure}[!h]
    \centering
    \includegraphics[scale=0.6]{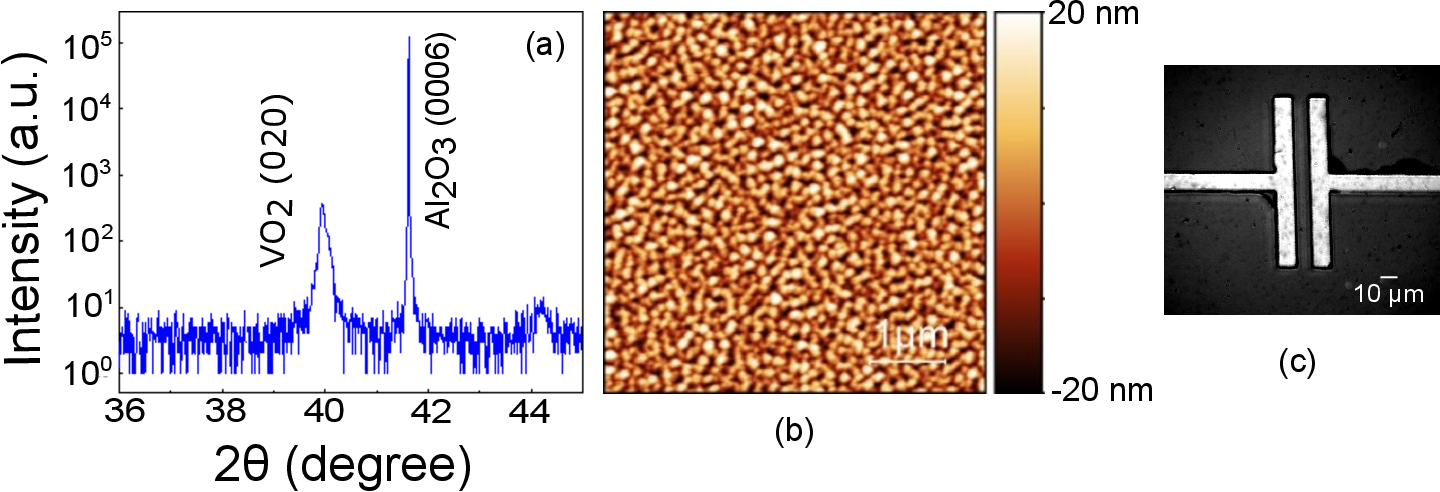}
    \caption{(a) XRD pattern of VO$_2$ on sapphire substrates. (b) Surface morphology AFM images (5x5 $\mu$m$^2$) of VO$_2$ films on sapphire substrates using PLD. (c) 10 $\mu$m × 100 $\mu$m device patterned on VO$_2$ film using optical lithography.}
    \label{fig:Samplefabrication}
\end{figure}

 The crystalline structure of the films was examined by X-ray diffraction (XRD), while the surface morphology was characterized using atomic force microscopy (AFM). As shown in Fig. \ref{fig:Samplefabrication}, the prominent diffraction peak corresponds to the Al$_2$O$_3$ (0006) reflection from the sapphire substrate. An additional diffraction peak observed at approximately 2$\theta$ = 39.9$^\circ$ is attributed to the (020) reflection of monoclinic VO$_2$, confirming the formation of the monoclinic VO$_2$ phase. The VO$_2$ thin films were patterned into device structures using standard optical lithography. A 50 nm thick Au layer was sputtered onto the lithographically defined regions to form the electrical contact electrodes. The resulting devices had a characteristic geometry of 10 $\mu$m × 100 $\mu$m [Fig. \ref{fig:Samplefabrication}(c)]. Electrical measurements were carried out using the Au electrodes as electrical contacts.

\section{Measurement setup}

Experimental setup for simultaneous electrical transport and optical imaging measurements: A green LED (521 nm) with Kohler illumination provides uniform illumination of the device through a series of lenses and an aperture. The reflected light from the sample is collected by a 20X objective (NA = 0.6) and directed through a 50:50 beam splitter to a monochrome camera via a tube lens for simultaneous imaging. The VO$_2$ device is mounted inside a transport-optical cryostat.\cite{Bajaj2026}

\begin{figure}[!h]
    \centering
    \includegraphics[scale=0.18]{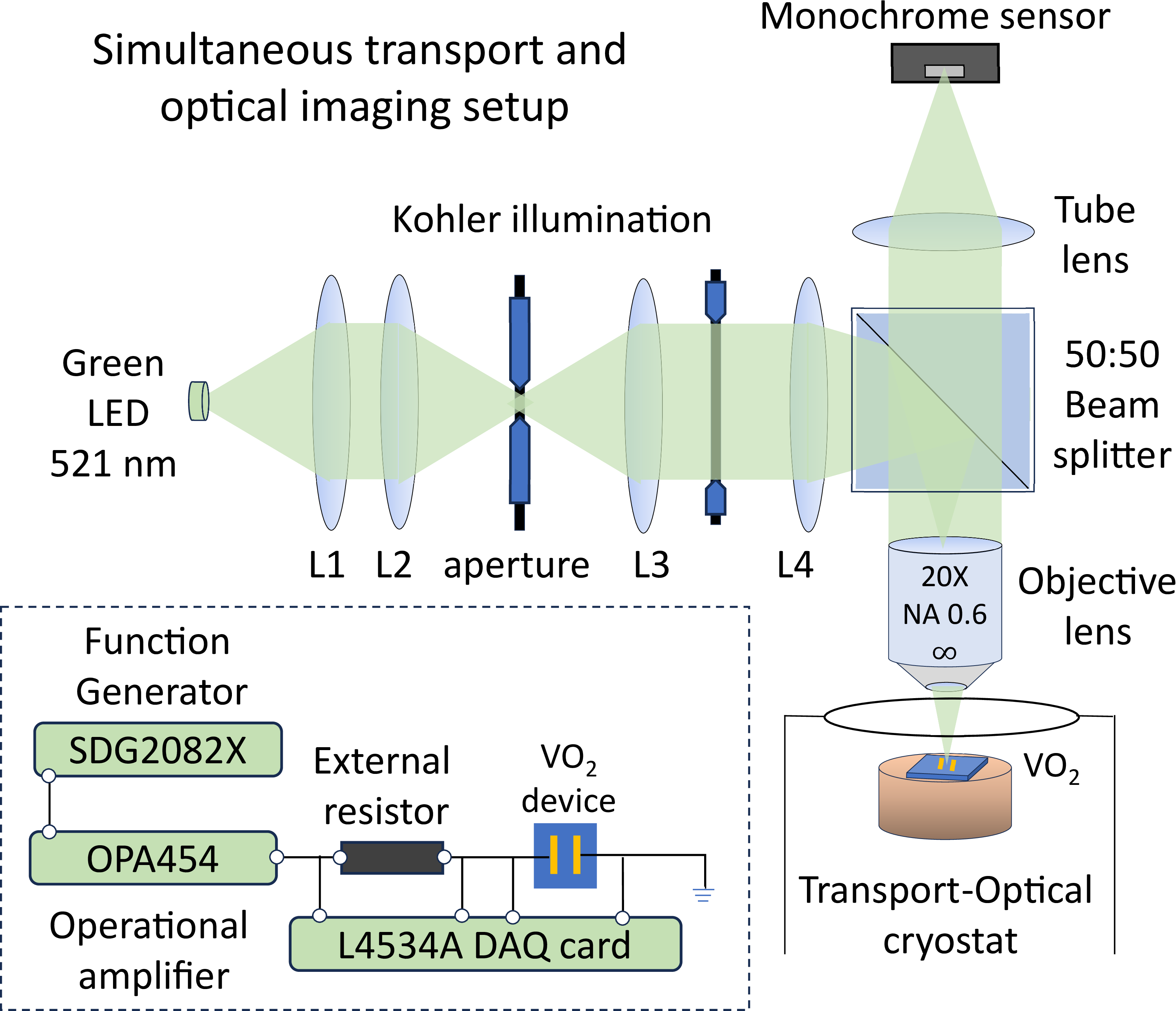}
    \caption{Schematic of the experimental setup for simultaneous electrical transport and optical imaging measurements.}
    \label{fig:measurement_setup}
\end{figure}

Electrical bias is applied using a function generator and an operational amplifier, with an external series resistor, while the current–voltage characteristics during electric breakdown are simultaneously recorded using a data acquisition card with high sampling rate. This configuration enables concurrent acquisition of I(V) data and optical visualization of filament formation during breakdown in the VO$_2$ device.

\section{Derivation of the Pre-Threshold Resistance Scaling Relation}

Under a constant applied voltage bias $V$, the 0D mean-field lump-circuit thermal balance equation is given in the main manuscript (Eq.~3) as:
\begin{equation}
\frac{\mathrm{d}T}{\mathrm{d}t} = \frac{1}{C_{\text{th}}} \left[ \frac{V^2}{R(T)} - K_{\text{th}}(T - T_0) \right] = 0
\implies
V^2 = K_{\text{th}} \Delta T R(T).
\label{EqS:SteadyState}
\end{equation}

where $\Delta T \equiv T - T_0$ is the temperature elevation above ambient temperature $T_0$.

In the sub-threshold pre-switching regime, we approximate the temperature-dependent device resistance via a Taylor expansion linearized about $T_0$ (for the negative temperature coefficient of resistance):
\begin{equation}
R(T) \approx R(T_0) - \left| \frac{\partial R}{\partial T} \right|_{T_0} \Delta T.
\label{EqS:LinearizedR}
\end{equation}
Thus, 
 \begin{equation}
\Delta T = \frac{R(T_0) - R(T)}{\left| \frac{\partial R}{\partial T} \right|_{T_0}}.
\label{EqS:DeltaT}
\end{equation}

Substituting Eq.~\ref{EqS:DeltaT} into the steady-state thermal power balance (Eq.~\ref{EqS:SteadyState}) yields:
\begin{equation}
V^2 = K_{\text{th}} \left( \frac{R(T_0) - R(T)}{\left| \frac{\partial R}{\partial T} \right|_{T_0}} \right) R(T) \implies R(T)^2 - R(T_0) R(T) + \frac{\left| \frac{\partial R}{\partial T} \right|_{T_0}}{K_{\text{th}}} V^2 = 0.
\label{EqS:QuadraticR}
\end{equation}

Now, the critical threshold voltage $V_c$ is defined in the main manuscript (Eq.~5) as:
\begin{equation}
V_c^2 = \frac{K_{\text{th}}}{4} \frac{R^2(T_0)}{\left| \frac{\partial R}{\partial T} \right|_{T_0}}.
\label{EqS:Vc_Def}
\end{equation}
Rearranging Eq.~\ref{EqS:Vc_Def} isolates the physical ratio:
\begin{equation}
\frac{\left| \frac{\partial R}{\partial T} \right|_{T_0}}{K_{\text{th}}} = \frac{R^2(T_0)}{4 V_c^2}.
\label{EqS:AlphaKth}
\end{equation}

Substituting Eq.~\ref{EqS:AlphaKth} back into the quadratic equation (Eq.~\ref{EqS:QuadraticR}):
\begin{equation}
R(T)^2 - R(T_0) R(T) + \frac{R^2(T_0)}{4} \left( \frac{V}{V_c} \right)^2 = 0.
\end{equation}
Dividing the entire quadratic equation by $R^2(T_0)$ casts it into normalized non-dimensional variables:
\begin{equation}
\left( \frac{R(T)}{R(T_0)} \right)^2 - \left( \frac{R(T)}{R(T_0)} \right) + \frac{1}{4} \left( \frac{V}{V_c} \right)^2 = 0.
\label{EqS:NormalizedQuadratic}
\end{equation}

Solving Eq.~\ref{EqS:NormalizedQuadratic} for the normalized resistance $\frac{R(T)}{R(T_0)}$:
\begin{equation}
\frac{R(T)}{R(T_0)} = \frac{1}{2} \pm \frac{1}{2}\sqrt{1 - \left(\frac{V}{V_c}\right)^2}.
\label{EqS:TheoreticalScaling}
\end{equation}

At the critical threshold ($V = V_c$), the discriminant vanishes, predicting a theoretical threshold resistance ratio:
\begin{equation}
\frac{R(T_c)}{R(T_0)} = \frac{1}{2}.
\label{EqS:TheoreticalRatio}
\end{equation}

In actual physical devices, higher-order non-linearities and non-ideal thermal dissipation cause the measured critical resistance ratio $\frac{R(T_c)}{R(T_0)}$ to deviate from $1/2$. To construct a universal empirical collapse curve while preserving the square-root functional form predicted by thermal runaway dynamics, we generalize the scaling relation as:
\begin{equation}
\frac{R(T)}{R(T_0)} = A \pm B \sqrt{1 - \left(\frac{V}{V_c}\right)^2}.
\label{EqS:GeneralScaling}
\end{equation}

Applying physical boundary conditions, Zero-bias condition ($V = 0$) and Critical switching threshold ($V = V_c$), determines constants $A = \frac{R(T_c)}{R(T_0)}$ and $B = 1 - \frac{R(T_c)}{R(T_0)}$ which modifies Eq.~\ref{EqS:GeneralScaling} to yield the final empirical scaling relation (Eq.~6 in the main manuscript):
\begin{equation}
\frac{R(T)}{R(T_0)} = \frac{R(T_c)}{R(T_0)} \pm \left[ 1 - \frac{R(T_c)}{R(T_0)} \right] \sqrt{1 - \left(\frac{V}{V_c}\right)^2}.
\label{EqS:FinalEmpiricalScaling}
\end{equation}

\section{Formulation of Moving-Boundary Dynamics of Filamentation}

To derive the moving-boundary dynamical equation governing filament width $a(t)$, we perform domain integration on the modified 2D reaction-diffusion equation given in the main manuscript (Eq.~12) across film thickness $d$ and transverse coordinate $x \in (-\infty, \infty)$:

\begin{equation}
d \int_{-\infty}^{\infty} \frac{\partial H(\hat{T})}{\partial t} \, \mathrm{d}x = \kappa d \int_{-\infty}^{\infty} \frac{\partial^2 \hat{T}}{\partial x^2} \, \mathrm{d}x + \int_{-\infty}^{\infty} d \, \frac{E^2}{\rho(\hat{T})} \, \mathrm{d}x - \int_{-\infty}^{\infty} h (\hat{T}(x) - T_0) \, \mathrm{d}x,
\label{EqS:1D_Integrated_PDE}
\end{equation}
evaluated using the collective coordinate temperature profile ansatz $\hat{T}(x; a(t))$ defined in the main text.

Integration of Individual Thermal Power Terms--

\textbf{Global Enthalpy Rate ($\mathcal{U}$)}: The total spatial enthalpy per unit length along the channel direction $z$ is defined as $\mathcal{U}(a) = \int_{-\infty}^{\infty} d \cdot H(\hat{T}(x)) \, \mathrm{d}x$. Splitting the domain into the metallic core ($|x| \le a$) and insulating tails ($|x| > a$):
\begin{equation}
\mathcal{U}(a) = d \int_{-a}^{a} (c_v \Delta + \varrho L_h) \, \mathrm{d}x + 2 d \int_{a}^{\infty} c_v \Delta \exp\left(-\frac{x - a}{\ell_{\text{th}}}\right) \mathrm{d}x.
\end{equation}

Evaluating the integrals:
\begin{align}
\text{Core Enthalpy:} \quad & d \int_{-a}^{a} (c_v \Delta + \varrho L_h) \, \mathrm{d}x = 2 a d (c_v \Delta + \varrho L_h), \\
\text{Tail Enthalpy:} \quad & 2 d c_v \Delta \int_{0}^{\infty} e^{-\frac{\xi}{\ell_{\text{th}}}} \, \mathrm{d}\xi = 2 d c_v \Delta \ell_{\text{th}} \quad (\text{where } \xi = x - a).
\end{align}

Summing these contributions yields the integrated spatial enthalpy:
\begin{equation}
\mathcal{U}(a) = 2 a d (c_v \Delta + \varrho L_h) + 2 d c_v \Delta \ell_{\text{th}}.
\end{equation}

For temporal dynamics of filament, using ($\frac{\mathrm{d}\mathcal{U}}{\mathrm{d}t} = \frac{\mathrm{d}\mathcal{U}}{\mathrm{d}a} \frac{\mathrm{d}a}{\mathrm{d}t}$) and  differentiate with respect to time:
\begin{equation}
\frac{\mathrm{d}\mathcal{U}}{\mathrm{d}t} = \frac{\mathrm{d}}{\mathrm{d}a}\left[ 2 a d (c_v \Delta + \varrho L_h) + 2 d c_v \Delta \ell_{\text{th}} \right] \frac{\mathrm{d}a}{\mathrm{d}t} = 2 d (c_v \Delta + \varrho L_h) \frac{\mathrm{d}a}{\mathrm{d}t} = \mathcal{E}_S \frac{\mathrm{d}a}{\mathrm{d}t}.
\label{EqS:dU_dt}
\end{equation}
The static tail term $2 d c_v \Delta \ell_{\text{th}}$ is independent of filament half-width $a(t)$ and drops out upon differentiation, showing that tail heat storage does not impede boundary velocity $\frac{\mathrm{d}a}{\mathrm{d}t}$.

\textbf{In-Plane Thermal Conduction}:
Integrating the lateral diffusion term across the infinite domain:
\begin{equation}
\kappa d \int_{-\infty}^{\infty} \frac{\partial^2 \hat{T}}{\partial x^2} \, \mathrm{d}x = \kappa d \left[ \frac{\partial \hat{T}}{\partial x} \right]_{-\infty}^{\infty} = 0,
\end{equation}
because temperature gradients vanish far from the filament boundary ($x \to \pm\infty$). Lateral conduction redistributes heat locally without acting as a global energy sink.

\textbf{Generated Joule Power ($\mathcal{P}_{\text{gen}}$)}:
Given the high conductivity contrast ($\sigma_M \gg \sigma_I$), current flow $I$ is localized to the metallic core ($|x| \le a$). With electric field $E = \frac{I}{2 a d \sigma_M}$, integrating generated power per unit length yields:
\begin{equation}
\mathcal{P}_{\text{gen}}(a, I) = \int_{-a}^{a} d \sigma_M E^2 \, \mathrm{d}x = 2 a d \sigma_M \left( \frac{I}{2 a d \sigma_M} \right)^2 = \frac{I^2 \rho_M}{2 a d}.
\end{equation}

\textbf{Substrate Cooling Loss ($\mathcal{P}_{\text{loss}}$):}
Integrating vertical interfacial dissipation across core and tail regions per unit length gives:
\begin{equation}
\mathcal{P}_{\text{loss}}(a) = h \left[ \int_{-a}^{a} \Delta \, \mathrm{d}x + 2 \Delta \int_{0}^{\infty} e^{-\frac{\xi}{\ell_{\text{th}}}} \, \mathrm{d}\xi \right] = 2 h \Delta (a + \ell_{\text{th}}).
\end{equation}

\textbf{Electrode-Gap Scaling and Circuit-Coupled Dynamics:}

Multiplying the power balance per unit length by electrode gap $L$ yields the global energy dynamic balance over the full 3D volume of the filament:
\begin{equation}
\mathcal{E}_S L \frac{\mathrm{d}a}{\mathrm{d}t} = \mathcal{P}(a, I) = \frac{I^2 \rho_M L}{2 a d} - 2 h \Delta (a + \ell_{\text{th}}) L.
\label{EqS:Current_Dynamics}
\end{equation}

with $\mathcal{E}_S=2 d (c_v \Delta + \varrho L_h)$.

\end{document}